\documentclass[preprint,showpacs,byrevtex,prl,10pt,twocolumn]{revtex4}%
\usepackage{amsfonts}
\usepackage{amsmath}
\usepackage{amssymb}
\usepackage[dvips]{graphicx}%
\providecommand{\U}[1]{\protect\rule{.1in}{.1in}}
\newtheorem{theorem}{Theorem}
\newtheorem{acknowledgement}[theorem]{Acknowledgement}

\begin{document}
\preprint{ }
\title{Thermalization Processes in Interacting Anderson Insulators}
\author{Z. Ovadyahu}
\affiliation{Racah Institute of Physics, The Hebrew University, Jerusalem 91904, Israel }

\pacs{72.80.Ng 73.61.Jc 72.20.Ee}

\begin{abstract}
This paper describes experiments utilizing a unique property of
electron-glasses to gain information on the fundamental nature of the
interacting Anderson-localized phase. The methodology is based on measuring
the energy absorbed by the electronic system from alternating electromagnetic
fields as function of their frequency. Experiments on three-dimensional (3D)
amorphous indium-oxide films suggest that, in the strongly localized \ regime,
the energy spectrum is discrete and inelastic electron-electron events are
strongly suppressed. These results imply that, at low temperatures, electron
thermalization \textit{and} finite conductivity depend on coupling to the
phonon bath. The situation is different for samples nearing the
metal-insulator transition; in insulating samples that are close to the
mobility-edge, energy absorption persists to much higher frequencies.
Comparing these results with previously studied 2D samples [Ovadyahu, Phys.
Rev. Lett., \textbf{108}, 156602 (2012)] demonstrates that the mean-level
spacing (on a single-particle basis) is not the only relevant scale in this
problem. The possibility of de-localization by many-body effects and the
relevance of a nearby mobility-edge (which may be a many-body edge) are discussed.

\end{abstract}
\maketitle

\section{Introduction}

The question of how Coulomb interactions affect Anderson localization has been
a challenging problem for decades. It was first addressed by Fleishman and
Anderson \cite{1} in the context of the stability of the insulating phase as
well as the mechanism of system thermalization and energy exchange involved in
hopping conductivity. Thermalization of Fermi-gas systems depends on inelastic
scatterings of electrons. Energy-exchange via electron-electron (e-e)
scattering is required for establishing the Fermi-Dirac thermal-distribution,
which defines the electron temperature. Such events lead to de-coherence of
the electrons and thus also control the quantum effects exhibited by the system.

The most frequently encountered mechanisms for electronic energy-transfer in
condensed matter systems are e-e and electron-phonon (e-ph) scatterings. The
latter is needed to maintain steady-state conditions when the system is driven
by an external source, and in particular, are responsible for the validity of
Ohm's law.

In metallic systems at sufficiently low temperatures the e-e inelastic-rate
$\gamma_{\text{in}}^{\text{e-e}}$ is usually the main source of scattering
\cite{2}. Both, $\gamma_{\text{in}}^{\text{e-e}}$, and the e-ph inelastic-rate
$\gamma_{\text{in}}^{\text{e-ph}}$ can be measured in the diffusive regime
based on weak-localization effects \cite{3}. While similar quantum effects
sometime extend into the hopping regime of the same system \cite{4}, a
theoretical framework to analyze such data and obtain inelastic scattering
rates is unfortunately not yet established. $\gamma_{\text{in}}^{\text{e-e}}$
in the hopping regime is defined in this work as the rate of energy-exchange
$\delta$E involved in electron-electron scatterings where $\delta$E$\neq$0.
These processes contribute to the \textit{overall} life-time broadening of the
electronic level, which is what our experiments are designed to capture. It
should be remarked that the energy-state broadening includes the contribution
of other mechanisms (such as electron-phonon inelastic scattering). In the
diffusive regime the contribution of e-e inelastic-rate may be separated from
the e-ph one as demonstrated by Bergmann \cite{3}. This is not yet achievable
by the methodology used here for the insulating regime. On the other hand,
this method yields information on a wide frequency range (rather than just an
average value for the inelastic-rate) and it can still be shown that relative
to the diffusive regime $\gamma_{\text{in}}^{\text{e-e}}$ is dramatically suppressed.

It was recently shown that in the two-dimensional (2D) hopping regime of
crystalline indium-oxide In$_{\text{2}}$O$_{\text{3-x}}$, $\gamma_{\text{in}%
}^{\text{e-e}}$ is suppressed by $\approx$six orders of magnitude relative to
its value at the diffusive regime at the same temperature \cite{5}. This was
based on utilizing a unique property of electron-glasses \cite{6}; using a
non-Ohmic field to take the system out of equilibrium, endows the system with
excess conductance that may be used as an empirical measure of the energy
absorbed by the electrons from the field. This technique allows a measurement
on systems with very small volume, it is sensitive enough to allow for weak
absorption from electric fields, and can be carried over a wide frequency range.

Here we report on measurements performed on Anderson-localized amorphous
indium-oxide films (In$_{\text{x}}$O) that exhibit three-dimensional (3D)
hopping transport. The results of our measurements suggest that e-e energy
exchange in In$_{\text{x}}$O is strongly suppressed relative to its value in
the diffusive regime at the same temperature. Analysis of these results
suggests that thermalization of the electronic system is governed by
$\gamma_{\text{in}}^{\text{e-ph}}$ as was the case in the 2D crystalline
version \cite{5}. However, approaching the metal-insulator transition by
reducing the quenched disorder, the perceived inelastic-rate tends towards the
$\gamma_{\text{in}}^{\text{e-e}}$ value typical of the diffusive regime. This
occurs while the system is still insulating; it exhibits
variable-range-hopping transport and its disorder is as strong as that of the
2D samples where $\gamma_{\text{in}}^{\text{e-e}}$ was highly suppressed
\cite{5}. We point out some similarity of these observations with a peculiar
temperature dependence of the conductivity in 3D systems near their
metal-insulator transition, which has been observed in several materials. The
role of dimensionality, inherent inhomogeneities, many-body effects, and other
issues that might be involved in bringing about an apparent de-localized
behavior are discussed.

\section{Experimental}

\subsection{Samples preparation and characterization}

Three batches of In$_{\text{x}}$O samples were used in this study. They were
prepared by e-gun evaporation onto room-temperature substrates using 99.999\%
pure In$_{\text{2}}$O$_{\text{3-x}}$ sputtering target pieces. Substrates were
either 1mm thick microscope glass-slides, or on 0.5$\mu$m SiO$_{\text{2}}$
layer thermally grown on
$<$%
100%
$>$
silicon wafers. Samples thickness d was 630\AA \ or 1050\AA \ for the
glass-slides, and d=750\AA \ for the Si wafers. Rate of deposition and
thickness were measured by a quartz thickness monitor calibrated using optical
interference measurements on thick MgF$_{\text{2}}$ films. Deposition was
carried out at the ambience of (1-3)$\cdot$10$^{\text{-4}}$ Torr oxygen
pressure maintained by leaking 99.9\% pure O$_{\text{2}}$ through a
needle-valve into the vacuum chamber (base pressure $\simeq$10$^{\text{-6}}$
Torr). Rates of deposition used for the samples reported here were typically
0.6-0.9~\AA /s. Under these conditions, the In$_{\text{x}}$O samples had
carrier-concentration n in the range (7-8)$\cdot$10$^{\text{19}}%
$cm$^{\text{\={-}3}}$ as measured by Hall-Effect at room temperatures on
samples that were patterned in a 6-probe configuration using stainless-steel
masks. These samples were prepared during the same deposition as the strips
used for the low temperature transport measurements. A standard Hall-bar
geometry was used with the active channel being a strip of 1 mm wide, and 10
mm long. The two pairs of voltage probes (that doubled as Hall-probes), were
spaced 3 mm from one another along the strip. This arrangement allowed us to
assess the large scale uniformity of the samples, both in terms the
longitudinal conductance and the Hall effect. Excellent uniformity was found
on these scales; resistivities of samples separated by 1~mm along the strip
were identical to within $\pm$5\%. No change (within the experimental error of
3\%) was observed in the hall effect due to annealing (tested for samples with
room temperature resistivity smaller than $\simeq$0.4$\Omega$cm which was the
highest $\rho$ in the samples studied in this work). On a mesoscopic scales
(10-100nm) however, In$_{\text{x}}$O films show compositional inhomogeneities;
the various effects these may have on transport properties of these films were
reported in \cite{7}.

As-deposited samples had room-temperature resistivity $\rho$ in excess of
10$^{\text{5}}\Omega$cm which, for the low temperature studies, had to be
reduced by several orders of magnitude. This was achieved by thermal annealing
at temperatures T$_{\text{a}}$%
$<$%
75 degree Celsius to prevent crystallization. For a comprehensive description
of the annealing process and the associated changes in the material
microstructure see \cite{7}.

\subsection{Measurements techniques}

Conductivity of the samples was measured using a two-terminal ac technique
employing a 1211-ITHACO current pre-amplifier and a PAR-124A lock-in
amplifier. Except when otherwise noted, measurements reported below were
performed with the samples immersed in liquid helium at T=4.1K maintained by a
100 liters storage-dewar. This allowed long term measurements of samples as
well as a convenient way to maintain a stable bath temperature. The ability to
keep the sample at $\approx$4K for long times is essential for these studies
where a typical series of measurements takes 4-6 weeks to accomplish. The ac
voltage-bias, used during the off-stress periods, was small enough to ensure
linear response conditions (judged by Ohm's law being obeyed within the
experimental error).

As measure of disorder we use the Ioffe-Regel dimensionless parameter,
k$_{\text{F}}\ell$=(9$\pi^{\text{4}}/$n)$^{\text{1/3}}\frac{\text{R}%
_{\text{Q}}}{\rho_{\text{RT}}}$ \ where R$_{\text{Q}}$=$\hslash/$e$^{\text{2}%
}$ is the resistance quantum. This is based on free-electron expressions using
the measured room-temperature resistivity $\rho_{\text{RT}}$ and the
carrier-concentration n, obtained from the Hall-Effect measurements, as parameters.

Several sources were used for exciting the system by non-Ohmic fields; the
internal oscillator of the PAR124A (up to 2 kHz and 10 Vrms) (Fluke PM5138A
(dc and up to 10 MHz and 40 Vpp), and Tabor WS8101 (up to 100 MHz and boosted,
when necessary, by Ophir 5084 RF power-amplifier). Complementary studies in
the microwave regime employed the high-power synthesizer HP8360B. Care was
taken in these experiments to use "RF-safe" components near the sample
immediate vicinity to minimize spurious heating. For the same reason, it was
ascertained, by performing four-probe measurements, that the contacts
resistance was always negligible relative to the sample resistance.

Optical excitation was accomplished by exposing the sample to AlGaAs diode
(operating at $\approx$0.88$\pm$0.05$\mu$m), placed $\approx$15mm from the
sample. The diode was energized by a computer-controlled Keithley 220
current-source. The samples were attached to a probe equipped with calibrated
Ge and Pt thermometers and were wired by triply-shielded cables to BNC
connectors at room temperatures. The effective capacitance of the wires was
$\leq$20pF. This allowed the use of 23-1500Hz ac technique without a
significant phase shift for any of the samples used here.

Fuller details of measurement techniques are given elsewhere \cite{8}.

\section{Results and discussion}

\subsection{Absorption measured via non-equilibrium transport}

The main technique used in this study is the `stress-protocol' previously used
in aging experiments \cite{9}. The procedure is composed of the following
stages (see Fig.1 for details): After the sample is equilibrated at the
measuring temperature (typically for 24 hours), its conductance versus time
~G(t) is recorded while keeping the electric field F$_{\text{0}}$ small enough
to be as close to the Ohmic regime as possible. This defines a baseline
`equilibrium G(0)'. Next, F is switched to a non-Ohmic "stress-field"
F$_{\text{stress}}$, which is kept on the sample for a time t$_{\text{w}}$.
Initially, F$_{\text{stress}}$ causes the conductance to increase by $\Delta
$G, but G is observed to keep increasing slowly throughout t$_{\text{w}}$.
Then the field is switched back to F$_{\text{0}}$ and the conductance is
continued to be measured for few thousand seconds. This last stage is depicted
in figure 1 as a relaxation of G(t) towards the equilibrium G(F$_{\text{0}}$)
with a logarithmic law characteristic of the relaxation processes in
electron-glasses \cite{10}. A measure of the magnitude of the excess
conductance that results from the stress is $\delta$G$_{\text{0}}$ (see inset
to Fig.1), defined by extrapolating the $\delta$G(t) curve to 1~second as
illustrated in the inset to Fig. 1. $\delta$G(t) is G(F$_{\text{0}}%
$,t)-G(F$_{\text{0}}$). The origin of time for the logarithmic plot in the
inset is taken as t$_{\text{w}}$+1 (i.e., 1 second after F$_{\text{stress}}$
is reset to the Ohmic field F$_{\text{0}}$).%

\begin{figure}[ptb]%
\centering
\includegraphics[
height=2.4976in,
width=3.3399in
]%
{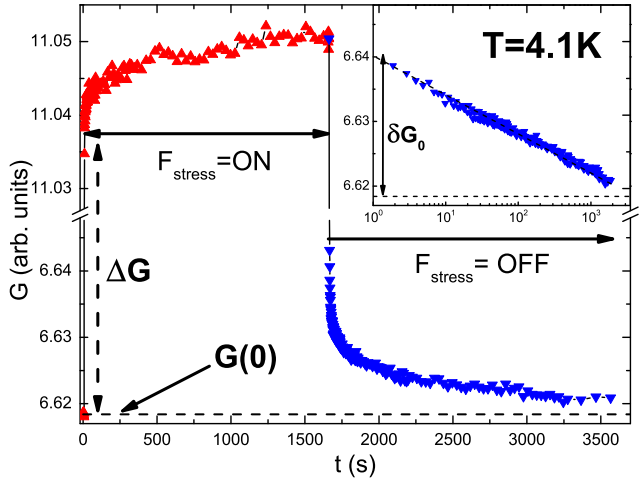}%
\caption{Conductance versus time G(t) illustrating a typical run of a
`stress-protocol'. The sample here is a In$_{\text{x}}$O with d=630\AA \ and
resistivity $\rho$=21.4$\Omega\cdot$cm at T=4.1K. The Ohmic and stress fields
here are F$_{\text{0}}$=100V/m and F$_{\text{stress}}$=10$^{\text{5}}$V/m,
both at 73Hz. The inset shows the logarithmic relaxation of $\delta$G(t) and
the definition of $\delta$G$_{\text{0}}$. Dashed lines delineate the
equilibrium conductance G(F$_{\text{0}}$).}%
\end{figure}

The relaxation of the excess conductance $\delta$G(t) is a manifestation of
the system approach to equilibrium from an excited state. A negative
time-derivative of $\delta$G(t) reflects energy release to the bath. This
energy may have been imparted to the system by a number of different
mechanisms; for example, exposing the sample to infrared radiation (Fig.2), or
changing its carrier-concentration using a nearby gate (Fig.3). In either
case, the ensuing logarithmic relaxation of $\delta$G$_{\text{0}}$ follows the
same relaxation law as that produced by the stress-field.%

\begin{figure}[ptb]%
\centering
\includegraphics[
height=2.5962in,
width=3.4255in
]%
{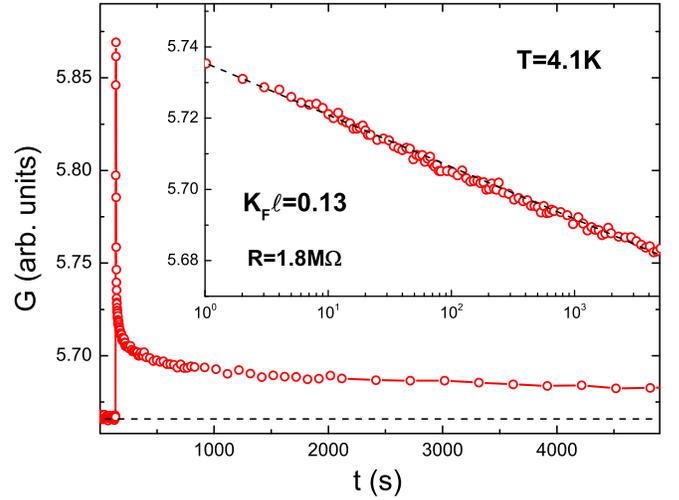}%
\caption{Excitation by IR exposure; after 12s from start of run, a small IR
emitter LED placed $\simeq$1 cm from the sample is energized by a 1mA current
for 3s then the LED is turned off. G(t), continuously monitored throughout the
run, reaches a peak ($\simeq$5.87 on the ordinate) then slowly decays towards
its equilibrium value. As in Fig.1, the ensuing relaxation is logarithmic
(inset). Note however, that the initial amplitude for the relaxation falls
short of the conductance peak (compare with the gate protocol in Fig.3). The
initial ($<$1s) fast decay is due to the heating that accompanies the IR
radiation. Dashed horizontal line marks the near-equilibrium G.}%
\end{figure}
%

\begin{figure}[ptb]%
\centering
\includegraphics[
height=2.6731in,
width=3.4255in
]%
{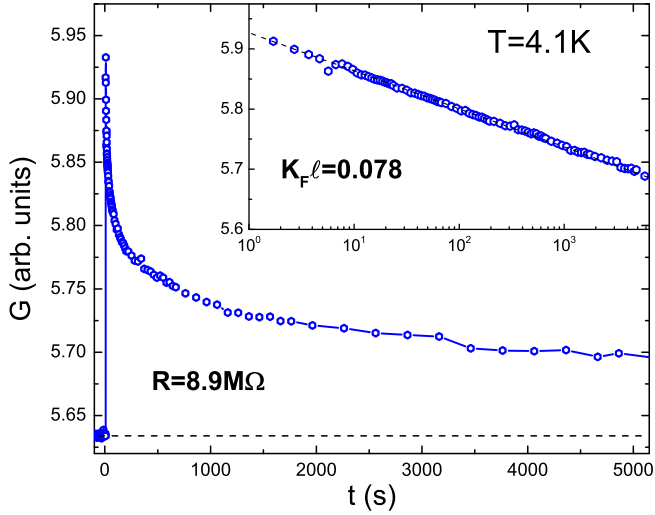}%
\caption{Exciting the sample by the `gate-protocol'; the sample is on a 1mm
thick glass substrate with a conducting silver-paint coating on its back side
acting as gate. After G(t) is monitored for t=90s to establish a baseline G,
the gate-voltage, initially at -198V is switched to +198V in 2s, and is held
there for the reminder of the run. The inset shows that the excess conductance
after the switch relaxes logarithmically. Note that the initial amplitude of
the relaxation process coincides essentially with the excitation peak.}%
\end{figure}

The mechanism by which stressing the system with a non-Ohmic field increases
the electronic energy is essentially Joule-heating; the energy absorbed by the
electrons gives rise to an excess phonons, making it somewhat `hotter' than
the bath. A steady-state may be established, while the stress-field is on, by
the flow of energy carried by the phonons in the sample into the thermal-bath.
The increased density of high energy phonons (over the phonon population in
equilibrium at the bath temperature), randomizes the charge configuration of
the electron-glass in a similar vein that raising the bath-temperature would
\cite{11}. This produces the excess conductance that relaxes back to its near
equilibrium value once the stress is relieved.

It is therefore plausible to take $\delta$G$_{\text{0}}$ as a measure of the
energy $\delta\varepsilon$ absorbed by the electronic system from the field.
As long as $\delta$G$_{\text{0}}$/G(0)$\ll$1, $\delta$G$_{\text{0}}$ is
arguably proportional to $\delta\varepsilon$.

It is emphasized that the only assumption we make in this procedure is that
$\delta\varepsilon$ enters the sample via the coupling of F$_{\text{stress}}$
to the\textit{ electronic system}. No other assumption is made. In particular,
the reason for the very sluggish release of this energy is not relevant for
our considerations in this work. The logarithmic nature of $\delta$G(t) is
taken as a convenient empirical fact that allows us to estimate the absorption
via transport measurements. This rationale was used in \cite{5} in the study
of the frequency dependence of the electronic absorption of two-dimensional
(2D) films of In$_{\text{2}}$O$_{\text{3-x}}$. In this study we extend the
study to three-dimensional (3D) films.

\subsection{ Electronic absorption versus frequency and disorder}

Our first goal here is to define a protocol that allows a meaningful
comparison between energy absorption from the stress-field applied to the
system at various frequencies. The natural choice is to normalize $\delta
$G$_{\text{0}}^{\text{f}}$ - the excess conductance measured under
F$_{\text{stress}}$ at a frequency f by $\delta$G$_{\text{0}}^{\text{dc}}$-
the excess conductance measured with a dc field while keeping the same
$\Delta$G/G(0) and the same t$_{\text{w}}$ for each tested frequency of the
applied stress. It was found that for f%
$<$%
30Hz $\delta$G$_{\text{0}}^{\text{f}}$ was indistinguishable (within the
experimental error) from $\delta$G$_{\text{0}}^{\text{dc}}$ and in the
experiments reported here we used the result for F$_{\text{stress}}$ operating
at f=11-23Hz as the normalizing value. For applying the stress-protocol at
frequencies above $\approx$1kHz, the conductance was measured at f=23Hz and
under low-bias conditions to ensure linear-regime measurement throughout the
protocol. A stress-field F$_{\text{stress}}$ with frequency f was capacitively
superimposed across the sample. The high frequency component of
F$_{\text{stress}}$ was filtered out in the input to the current preamplifier
(in addition to the band-pass filtering in the 124A lock-in amplifier) such
that the conductance was measured at the low frequency. The amplitude of
F$_{\text{stress}}$ was adjusted to achieve the desired $\Delta$G/G(0) based
on the conductance reading at f=23Hz. The relaxation part of the protocol was
always measured under near-Ohmic conditions and at a low frequency (or at dc).

Results of absorption at different frequencies for three of the 3D samples
studied with the protocol described above are shown in Fig.4. For comparison,
the figure includes the 2D samples studied previously \cite{5}.%
\begin{figure}[ptb]%
\centering
\includegraphics[
height=3.7524in,
width=3.4255in
]%
{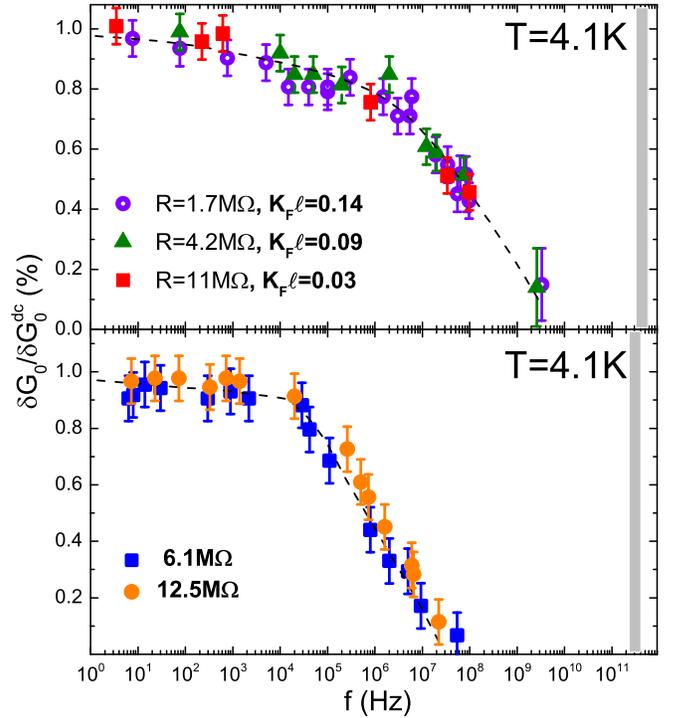}%
\caption{Relative absorption as function of the stress-field frequency for
three In$_{\text{x}}$O samples (d=630\AA \ for circles and triangles, and
d=1050\AA \ for the squares). These are labeled by their resistance at 4.1K
and their k$_{\text{F}}\ell$ values\ (top plate). The larger error-bars for
the f$>$2GHz data are based on F$_{\text{stress}}$ in the microwaves range
where $\Delta$G/G(0) was typically limited to $\simeq$0.1 rather than the 0.6
for the smaller frequencies. These results are compared with the data obtained
on 2D In$_{\text{2}}$O$_{\text{3-x}}$ films (d=52\AA , taken from \cite{5}).
The shaded area mark the respective ranges for $\gamma_{\text{in}}%
^{\text{e-e}}$ of these materials in their diffusive regime at $\simeq$4K
(based on reference \cite{7} and reference \cite{5} for In$_{\text{x}}$O and
In$_{\text{2}}$O$_{\text{3-x}}$ respectively). Dashed lines are guides for the
eye.}%
\end{figure}

Looking at these data one notes the following:

\begin{itemize}
\item[1] The general trend is for the absorption to decrease with frequency,
and in all three cases shown in Fig.4 there is a faster decline of the
absorption with f above a certain roll-off frequency f$_{\text{RO}}$. It is
observed that f$_{\text{RO}}$ is considerably lower than the respective
electron-electron inelastic rate $\gamma_{\text{in}}^{\text{e-e}}$ of the
material (based on measurements performed on a 3D In$_{\text{x}}$O system in
its diffusive regime at the same temperature).

\item[2] The frequency range over which the absorption decays appears to be
rather wide, extending over several decades. For comparing results under
different conditions, we take f$_{\text{RO}}$ to be the frequency where
$\delta$G$_{\text{0}}^{\text{f}}$/$\delta$G$_{\text{0}}^{\text{dc}}$=1/2.

\item[3] For the range of disorder shown, there seems to be no dependence on
the disorder in either 3D or 2D; samples with different disorder show
essentially identical absorption versus f curves. However, as will soon
transpire, this is only true for samples in the strongly-localized regime.
\end{itemize}

It has been shown \cite{5} that the roll-off frequency in the two-dimensional
In$_{\text{2}}$O$_{\text{3-x}}$ samples (lower graph in Fig.4) is consistent
with the electron-phonon inelastic scattering-rate $\gamma_{\text{in}%
}^{\text{e-ph}}$ of the material at 4K. This estimate was based on the
assumption that, while under a dc stress-field, the system reaches
steady-state conditions in which case the energy absorbed by the electrons
from F$_{\text{stress}}$ equals the energy dissipated into the bath. This may
be described by the following expression:%
\begin{equation}
\frac{\text{V}^{\text{2}}}{\text{R(V)}}\text{=C}_{\text{el}}\text{(T*)}%
\centerdot\text{U}\centerdot\Delta\text{T}\centerdot\gamma\text{(T*)} \tag{1}%
\end{equation}

The L.H.S of Eq.1 is the Joule-heating term; V is the voltage across the
sample, and R(V) is its resistance under V. The R.H.S is the heat-removal rate
while F$_{\text{stress}}$ is on, and assuming steady-state conditions. In this
equation C$_{\text{el}}$ is the electronic heat-capacity, U the sample volume,
T* is an "effective electron-temperature", $\Delta$T is T*-T$_{\text{bath}}$
(the bath temperature). For our samples being macroscopic, $\gamma$(T*) should
coincide with the inelastic electron-phonon rate $\gamma_{\text{in}%
}^{\text{e-ph}}$.

The parameters needed for calculating, $\gamma$(T*) are all obtained from
measurements on the respective sample except for C$_{\text{el}}$ and T*.
Following the procedure used in \cite{12}, T* may be estimated from G(T) data
(the uncertainty associated with this procedure will be commented on below).

The electronic heat capacity C$_{\text{el}}$ is$~$proportional to the
temperature and to $\partial$n$/\partial\mu$, the thermodynamic
density-of-states of the material. Since the carrier-concentration of the
In$_{\text{x}}$O samples used in this work is comparable to that of
In$_{\text{2}}$O$_{\text{3-x}}$ (and therefore C$_{\text{el}}$ for them should
be similar), we can estimate of the ratio between $\gamma_{\text{in}%
}^{\text{e-ph}}$ of In$_{\text{x}}$O to that of the In$_{\text{2}}%
$O$_{\text{3-x}}$ samples by:%

\begin{equation}
\frac{\gamma_{\text{a}}\text{(T*)}}{\gamma_{\text{c}}\text{(T*)}}\text{=}%
\frac{\text{V}_{\text{a}}^{\text{2}}\centerdot\text{R}_{\text{c}}%
\text{(V}_{\text{c}}\text{)}\centerdot\text{U}_{\text{c}}\centerdot
\Delta\text{T}_{\text{c}}}{\text{V}_{\text{c}}^{\text{2}}\centerdot
\text{R}_{\text{a}}\text{(V}_{\text{a}}\text{)}\centerdot\text{U}_{\text{a}%
}\centerdot\Delta\text{T}_{\text{a}}} \tag{2}%
\end{equation}
where the subscripts `a' and `c' signify values for the amorphous and
crystalline material respectively. The voltages used in the respective
experiments are extracted from the conductance-voltage G(V) measurements of
the respective samples. For the samples we chose for the analysis below, these
data are shown in Fig.5:%

\begin{figure}[ptb]%
\centering
\includegraphics[
height=2.6809in,
width=3.339in
]%
{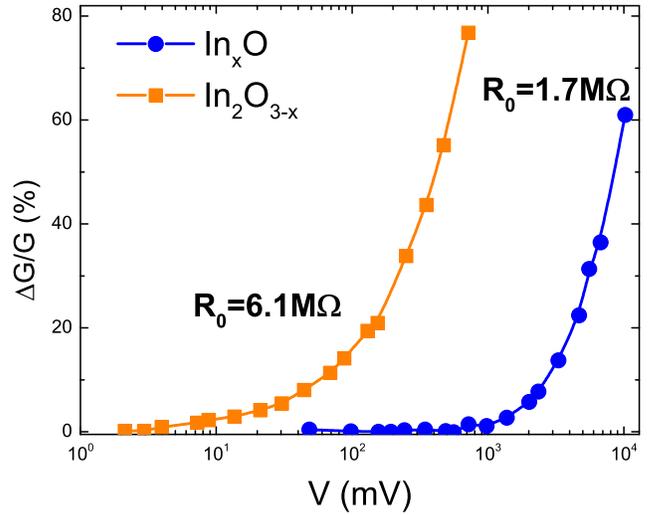}%
\caption{The fractional change of conductance due to applied voltage. The data
are for a 2D In$_{\text{2}}$O$_{\text{3-x}}$ film (squares, same sample as in
Fig.4 bottom plate), and a 3D In$_{\text{x}}$O sample (circles, same as in
Fig.4 top plate).}%
\end{figure}

The conditions that were used in the stress-protocol for our experiments were
$\Delta$G(0)/G(0)$\approx$0.6, and the associated F$_{\text{stress}}$ was
applied for t$_{\text{w}}$=1400~ seconds. To achieve this $\Delta$G(0)/G(0) a
voltage of 10.2V=V$_{\text{a}}$ was needed for the In$_{\text{x}}$O sample as
compared to 0.5V=V$_{\text{c}}$ for the crystalline sample \cite{5}, as can be
read from Fig.5. Note that these values (with the respective resistances),
mean that the power invested from the (dc) field is $\approx$1500-times larger
for the In$_{\text{x}}$O sample. The volumes for these samples were:
10$^{\text{-14}}$m$^{\text{3}}$=U$_{\text{c}}$ and 1.2$\cdot$10$^{\text{-13}}%
$m$^{\text{3}}$=U$_{\text{a}}$. The \textquotedblleft effective
temperature\textquotedblright\ T$^{\text{*}}$ for the In$_{\text{2}}%
$O$_{\text{3-x}}$ sample estimated in \cite{5} was 4.8K$\approx$T$_{\text{c}%
}^{\text{*}}$. This was based on the measured G(T) data for the sample under
steady-state conditions. Using the same logic here, the respective
T$^{\text{*}}$for the In$_{\text{x}}$O sample, based on its G(T) data, is
5.3K$\approx$T$_{\text{a}}^{\text{*}}$. The conductance versus temperature
curves for this sample and two other samples of the same In$_{\text{x}}$O
batch with different degrees of disorder are shown in Fig.6 below.%

\begin{figure}[ptb]%
\centering
\includegraphics[
height=2.5789in,
width=3.4255in
]%
{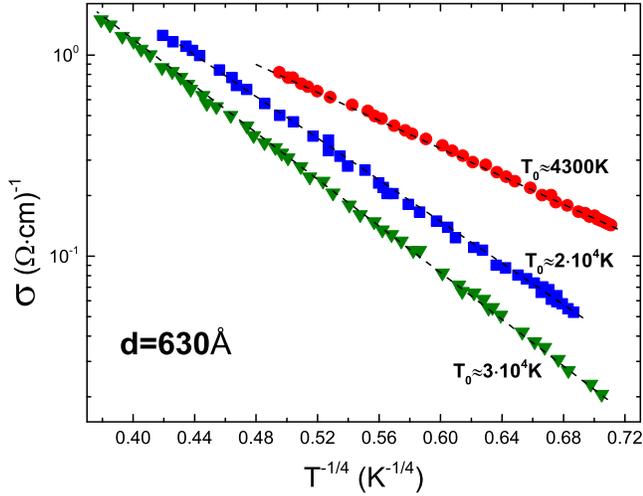}%
\caption{The temperature dependence of conductivity $\sigma$ for three
In$_{\text{x}}$O samples with different degrees of disorder (obtained by
thermal annealing from a single deposition batch). Samples are labeled with
their characteristic energy T$_{\text{0}}$ extracted from the data plotted to
conform with: $\sigma$(T)=$\sigma$(0)$\cdot$exp$\left(  \text{-[T}_{\text{0}%
}\text{/T]}^{\text{1/4}}\right)  $ }%
\end{figure}
It should be remarked that using the sample resistance as a thermometer to
obtain T* is a dubious procedure; in general, the resistance change due to the
applied non-Ohmic field is not \textit{just} a heating effect. In fact, deep
in the hopping regime \textit{most} of the resistance decrease is actually due
to field-assisted hopping \cite{13}. Hopping conductivity may be significantly
enhanced by field while a negligible increase of the electrons "temperature"
is observed; application of microwaves for example, may cause a $\Delta$G%
$>$%
0 with essentially no-heating \cite{14}, and the data in Fig.4 demonstrate
that this adiabatic effect gradually manifests itself even at lower
frequencies. In using Eq.2 for comparing between $\gamma_{\text{a}}$ and
$\gamma_{\text{c}}$ it is tacitly assumed that the part (reflected in $\Delta
$G) due to Joule-heating is the same for both materials. The error in the
determination of T$^{\text{*}}$ by assuming that G is a faithful thermometer,
is partially reduced in the ratio $\Delta$T$_{\text{c}}$/$\Delta$T$_{\text{a}%
}$.

With these reservations, we use the above parameters into Eq.2 which gives:
$\gamma_{\text{a}}$/$\gamma_{\text{c}}$=65. This value is not far from the
factor of $\approx$75 difference between the roll-off frequencies of the two
materials (see Fig.4), which in view of the uncertainties inherent in the
procedure should be judged as a satisfactory agreement.

The larger $\gamma_{\text{in}}^{\text{e-ph}}$ of the amorphous material
relative to In$_{\text{2}}$O$_{\text{3-x}}$ found in this work is plausible;
$\gamma_{\text{in}}^{\text{e-ph}}\approx$10$^{\text{8}}~\sec^{\text{-1}}$at
T$\approx$4K is typical of many degenerate Fermi systems \cite{15}. As a rule,
phonons in amorphous materials are softer than the crystalline version of the
same substance. Soft phonon-modes of the amorphous phase usually lead to
enhanced electron-phonon coupling. This is probably part of the reason why
In$_{\text{x}}$O, with less disorder and high carrier-concentration, is a
superconductor below few degrees Kelvin while In$_{\text{2}}$O$_{\text{3-x}}$
remains a normal conductor down to $\approx$12mK even when in the metallic
regime \cite{16}.

As to the second issue listed above; the extended range over which the
absorption decreases with frequency, especially conspicuous for the
In$_{\text{x}}$O samples. This wide range of $\gamma_{\text{in}}^{\text{e-ph}%
}$, is presumably a consequence of a wide distribution of localization-lengths
$\xi$,$~$an inherent property of the disordered system. It is noteworthy that
the absorption starts to diminish quickly at a rather small frequency (Fig.4)
suggesting a reduced $\langle\gamma_{\text{in}}^{\text{e-ph}}\rangle$. This is
not surprising; note that each inelastic scattering of electron by a phonon
involves the overlap of an initial state
$\vert$%
\textit{i}$\rangle$ with a final state
$\vert$%
\textit{j}$\rangle$ and a phonon with the energy that matches the energy
difference between these sites. Unless compensated somehow by a
disorder-enhanced phonon-electron matrix element \cite{17}, this inevitably
should lead to reduced transition rates relative to the diffusive regime where
both states are extended. Unfortunately, there is yet no theory for
$\gamma_{\text{in}}^{\text{e-ph}}$ in the hopping regime to compare our
results with.

\subsection{Reduced decoherence in the insulating regime?}

The frequency dependence of the absorption results discussed above suggests
that $\gamma_{\text{in}}^{\text{e-e}}$ (and, in some sense, also
$\gamma_{\text{in}}^{\text{e-ph}}$) is suppressed relative to its typical
values in the diffusive regime. In the case of $\gamma_{\text{in}}%
^{\text{e-e}}$ this reduction is by several orders of magnitude (see Fig.4).
These rate being the main sources of decoherence in most disordered electronic
systems the question arises: Does it also mean that there is less decoherence
in the Anderson insulator phase?

If one ignores the possibility that the highly disordered phase may breed
decoherence agents that were not present in the diffusive regime, like new
types of two-level-systems or local magnetic moments \cite{18}, then the
answer is yes; the insulating phase may have reduced inelastic rate relative
to the diffusive phase. This however applies to the coherence \textit{time};
the \textit{spatial} extent of the electron coherence will be limited by the
highly reduced diffusion constant. There is experimental evidence for quantum
coherent effects in the Anderson localized regime that, in some respects, are
more prominent than in the diffusive regime but the phase-coherent length is
limited to the hopping-length \cite{19}. The most compelling evidence for
quantum-coherent effects is the anisotropy of conductance-fluctuations
produced by magnetic field at different orientations is mesoscopic samples as
well as in the magnetoresistance of two-dimensional samples \cite{4}. The
existence of quantum-coherent effects deep in the hopping regime ought not be
surprising. The overlap between the initial and final state is affected by the
interference between different spatial trajectories. The phonon involved in
the actual transition does not destroy the interference once the phonon
wavelength exceeds the hopping-length, which inevitably happens at
sufficiently low temperatures \cite{4}.

The suppression of $\gamma_{\text{in}}^{\text{e-e}}$ may appear as a natural
consequence of the spectrum discreteness associated with localization
\cite{20,21}. On closer examination, and taking into account the role of
Coulomb interaction, this issue is more complicated and yet unresolved. The
problem was first raised by Fleishman and Anderson \cite{1}. They considered
several scenarios by which interactions may modify the single-particle aspects
of Anderson-localized systems, while noting that G(T) of these systems still
conforms to variable-range-hopping law. It may be illuminating then to see
what we can infer from the VRH conductivity of the samples on the spectrum discreteness.

The space-confinement due to localization forces a discrete energy spectrum
with a mean level-spacing of order (many-body effects may modify this
expression \cite{20}):%

\begin{equation}
\delta\text{(}\xi\text{)}\approx\left(  \partial\text{n}/\partial\mu\xi
^{3}\right)  ^{\text{-1}}\tag{3}%
\end{equation}

To get an estimate of $\delta$($\xi$) one may use data for the temperature
dependence of the hopping conductivity. This version of In$_{\text{x}}$O obeys
Mott's law (c.f., Fig.5) $\sigma(T)$=$\sigma(0)\centerdot\exp\left[
\text{-}\left(  \text{T}_{\text{0}}\text{/T}\right)  ^{\text{1/4}}\right]  $
where T$_{\text{0}}$ is given by \cite{22}:%

\begin{equation}
\text{k}_{\text{B}}\text{T}_{\text{0}}\approx\left(  \frac{\text{3}}%
{\partial\text{n}/\partial\mu\xi^{3}}\right)  \tag{4}%
\end{equation}

Using the data for $\sigma$(T) (Fig.6) and Eq.4, the mean-level-spacing for
the two samples studied in Fig.4 is $\delta$($\xi$)$\approx$10$^{\text{4}}%
$K$\ggg$4K, and it is also much larger than E$_{\text{C}}\approx$e$^{\text{2}%
}$/($\kappa\xi$) where $\kappa\simeq$10 is the dielectric constant for
indium-oxide \cite{23}, and $\xi\geq$10\AA \ as will be shown below.
Therefore, unless many-body physics plays a role, it would appear that the
spectrum discreteness is large enough to suppress electron-electron inelastic
scattering. Thermalization of the electronic system and non-zero conductivity
then depend on the existence of a continuous bath, presumably phonons.
Fleishman and Anderson \cite{1} reached this conclusion for the localized
system in the limit of short range interaction. Our absorption versus
frequency results (Fig.4) are consistent with their conclusions for a
realistic interaction range, probably of the order of the hopping-length
r(T)$\approx\xi$(T$_{\text{0}}$/T)$^{\text{1/3}}$ or $\approx\xi$%
(T$_{\text{0}}$/T)$^{\text{1/4}}$ in 2D or 3D respectively

The suppression of $\gamma_{\text{in}}^{\text{e-e}}$ of Anderson insulators
turns out however to be true only for samples in the strongly disordered
regime; things appear to be more complicated as the MIT is approached.

The frequency dependence of the absorption measured on two samples that were
annealed to bring their k$_{\text{F}}\ell$ to 0.19 and 0.21, is shown in
Fig.7. These data exhibit a different trend than the more disordered samples
shown in Fig.4 above; the absorption still decreases with frequency but it
seems to extend to higher frequencies, possibly even surpassing the
$\gamma_{\text{in}}^{\text{e-e}}$ of the diffusive regime at this temperature.
At the same time, these samples are on the insulating side of the MIT, and
their mean-level-spacing is still much larger than both E$_{\text{C}}$ and the
bath temperature; their characteristic energies T$_{\text{0}}$ (defined by the
VRH conductivity expression) are $\approx$4300K (Fig.6) and $\approx$6600K
(Fig.9 below) for the sample with k$_{\text{F}}\ell$=0.21 and 0.19 respectively.%

\begin{figure}[ptb]%
\centering
\includegraphics[
height=2.4163in,
width=3.4255in
]%
{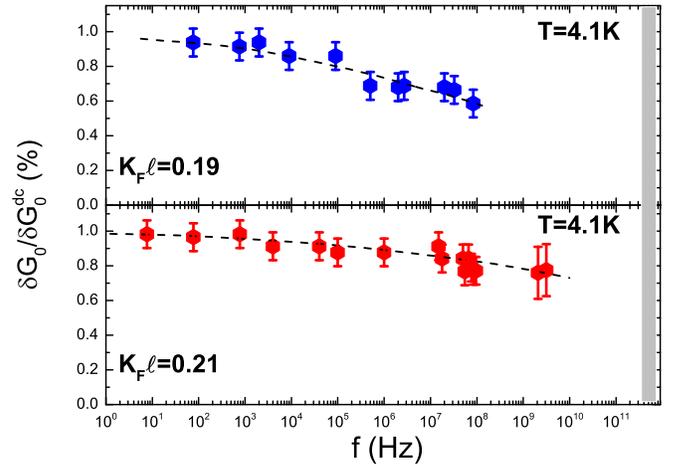}%
\caption{Relative absorption as function of the stress-field frequency for the
two In$_{\text{x}}$O samples that are nearest to the MIT (labeled by their
k$_{\text{F}}\ell$ values). The shaded area mark the\ typical range of
$\gamma_{\text{in}}^{\text{e-e}}$ of In$_{\text{x}}$O in its diffusive regime
at $\simeq$4K (based on reference \cite{7}). Dashed lines are guides for the
eye. Note the difference between these results and the results for the more
disordered samples (Fig.4 top plate), }%
\end{figure}

More intriguing is the observation that the disorder of the samples in Fig.4
is considerably larger than that of the 2D samples \{Fig.3 in \cite{5} and
Fig.4\}; the disorder in the latter is comparable to that of these 3D samples
in terms of $\delta$($\xi$) (and in terms of bulk resistivity the disorder in
the 2D films it is even \textit{much} \textit{weaker} than in the 3D samples).
If the reason for the energy-absorption cut-off in the 2D samples is the
spectrum-discreteness of the electronic states (having $\delta$($\xi$)$\gg
$k$_{\text{B}}$T) then why the same is not sufficient for 3D samples that
exhibit equally large $\delta$($\xi$)?

\subsection{The "T* problem"}

A possible direction to look for the difference between the 2D and 3D results
may be related to the materials; as noted above amorphous and crystalline
version of indium-oxide have their differences but none that seems relevant
for this particular feature. On the other hand, there is reason to suspect
that \textit{dimensionality} plays a role: Just insulating 3D samples of
In$_{\text{2}}$O$_{\text{3-x}}$ exhibit transport peculiarities that are not
observed in 2D \cite{24}. The 3D samples exhibited insulating characteristics
only when cooled below a disorder-dependent T*. Above this temperature, they
show a metallic-like G(T) law. On the other hand, 2D samples of this material,
with the same range of bulk resistivity exhibited insulating behavior
($\sigma\rightarrow$0 when T$\rightarrow$0) over the same temperature range
\cite{24}. Three dimensional samples of In$_{\text{x}}$O also exhibited the
same "T*-problem" \cite{24}. To illustrate, an example of the "T*-problem" is
shown in Fig.8 using one of the currently studied specimen. Note that, at low
temperatures, G(T) exhibits VRH conductivity (inset to Fig.8). This implies
G(T$\rightarrow$0)=0, which means the system is \textit{insulating}. However
above T*$\approx$60K the conductance law changes, and if one has no knowledge
of the behavior of G(T) at lower temperatures one will conclude, extrapolating
along the dashed curve in the main figure that G(T$\rightarrow$0)%
$>$%
0, that the system is actually a \textit{metal}.

A similar G(T) anomaly appears in quite a few other 3D systems (note that this
feature is easier to identify when G(T) rather than R(T) is plotted). Such a
peculiar G(T) may be seen in a series of amorphous Mn$_{\text{x}}%
$Si$_{\text{1-x}}$ samples \cite{25}, in amorphous Si$_{\text{1-x}}%
$Cr$_{\text{x}}$ \cite{26}, in GeAl \cite{27}, in granular aluminum \cite{24},
in crystalline GeSb$_{\text{2}}$Te$_{\text{4}}$ \cite{28}, and in
a-Gd$_{\text{x}}$Si$_{\text{1-x}}$ samples \cite{29}. We are not aware of any
3D system that was tested over a wide temperature range near its MIT without
showing the ambiguous G(T) characterizing the T*-problem. This feature may be
generic to 3D systems near their Anderson transition.%

\begin{figure}[ptb]%
\centering
\includegraphics[
height=2.5434in,
width=3.4255in
]%
{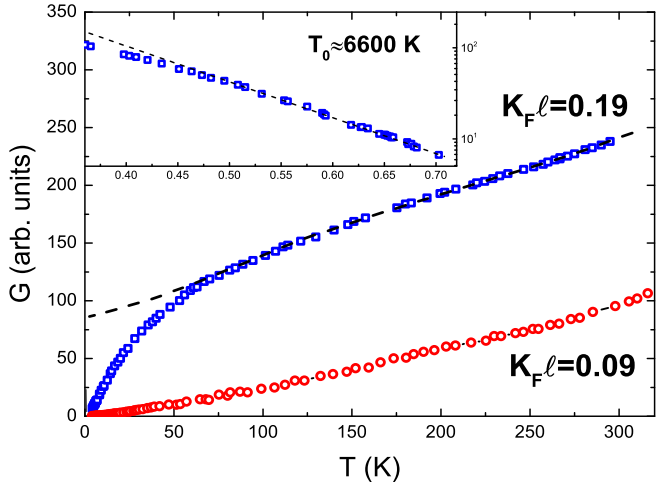}%
\caption{Conductance versus temperature for two of the samples studied for
absorption \ as function of frequency The sample with k$_{\text{F}}\ell$=0.19
illustrates the `T*-problem' - its G(T) data for T$>$60K extrapolates to a
finite G at T=0 (dashed line). The inset shows that the low temperature
dependence of this sample is variable range hopping. Compare these G(T) data
with the data in \cite{25} Fig.1b and with reference \cite{24} Fig.14.}%
\end{figure}

The phenomenology associated with this anomalous G(T), in particular the
systematic increase of T* with disorder, and the absence of the effect in 2D
\cite{24}, raised the possibility that T* signifies the mobility-edge; a
threshold energy E$_{\text{C}}$ separating extended states for E%
$>$%
E$_{\text{C}}$ from localized states for E%
$<$%
E$_{\text{C}}$. This indeed might account for the experimental observations.
However, the values for T* necessary for this line of explanation turned out
to be smaller than what one (perhaps naively) anticipates. Note that near the
metal-insulator transition the distance $\Delta$E to the mobility-edge is
expected to obey \cite{30}:%

\begin{equation}
\Delta\text{E}\equiv\text{%
$\vert$%
E}_{\text{C}}\text{-E}_{\text{F}}\text{%
$\vert$%
=E*%
$\vert$%
(g-g}_{_{\text{C}}}\text{)/g}_{_{\text{C}}}\text{%
$\vert$%
}^{\nu} \tag{5}%
\end{equation}
where, E$_{\text{F}}$ is the Fermi-energy, g is the dimensionless conductance,
g$_{_{C}}$ is the dimensional-conductance value at the MIT transition, and the
exponent $\nu$ is $\approx$1. To fit the G(T) data to Eq.5 it was necessary to
use for E*, a value considerably smaller than the Fermi energy of the
material, which shed some doubts on the notion that T* reflects the
mobility-edge \cite{24}.

The current results, and in particular the apparent role of dimensionality,
instigated a renewed look at these phenomena. The absorption experiments
(Fig.7) and the G(T) behavior of just insulating 3D samples have this in
common: Both exhibit diffusive characteristics, a tendency that becomes more
conspicuous as they further approach the MIT, and both involve probing the
system away from its ground-state: The stress experiments take the system far
from equilibrium, the T* problem is a finite temperature phenomenon. Indeed, a
simple explanation of the G(T) anomaly might be related to a
temperature-dependent probing length. An insulating sample will exhibit a
diffusive G(T) law when, for example, L$_{\text{in}}$%
$<$%
$\xi$ where L$_{\text{in}}$=L$_{\text{in}}$(T) is the inelastic diffusion
length. This however cannot account for the experimental T*-problem unless one
assumes either, an unusual energy dependence for $\xi$ or a specific $\xi$ -
distribution \cite{31}. A many-body scenario may have to be considered.

Let us examine the T*-problem in the context of the current issue assuming for
the moment it is in fact a mobility-edge. If E$_{\text{C}}$ is not far above
E$_{\text{F}}$, then new avenues for electron-electron energy exchange may
become available. It is thus of interest to find out how close is a system
with a given k$_{\text{F}}\ell$ to the transition. This may be estimated from
the dependence of the localization-length $\xi$ on the order-parameter
k$_{\text{F}}\ell$ of the sample in question. The localization length is
evaluated using the G(T) data with Eq.4, this yields the $\xi$(k$_{\text{F}%
}\ell$ ) plot shown in Fig.9:%

\begin{figure}[ptb]%
\centering
\includegraphics[
height=2.6377in,
width=3.3399in
]%
{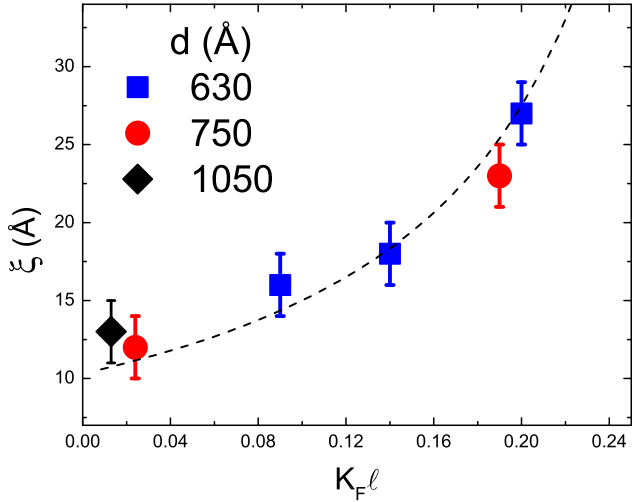}%
\caption{The dependence of the localization length $\xi$ on the Ioffe-Regel
parameter k$_{\text{F}}\ell$ for several of the studied films.}%
\end{figure}

The dependence of $\xi$ on k$_{\text{F}}\ell$ in this figure fits reasonably
well the expression: $\xi$=$~\frac{\xi_{\text{0}}}{\text{(k}_{\text{F}}%
\ell\text{)}_{_{\text{C}}}}\left[  \text{1-$\frac{\text{k}_{\text{F}}\ell
}{\text{(k}_{\text{F}}\ell\text{)}_{_{\text{C}}}}$}\right]  ^{\text{-1}}$
which is a variation on on the scaling form \cite{30} (with $\nu$=1):%

\begin{equation}
\xi=\xi_{\text{0}}\text{%
$\vert$%
(g-g}_{_{\text{C}}}\text{)/g}_{_{\text{C}}}\text{%
$\vert$%
}^{-\nu} \tag{6a}%
\end{equation}
where $\xi_{\text{0}}$ is the value of the localization-length far from the
MIT. Fitting the dependence of $\xi$ on k$_{\text{F}}\ell$ (Fig.9) yields
$\xi_{\text{0}}$=11\AA ,\ which is of the order of the Bohr-radius,
a$_{\text{B}}\simeq$15\AA \ for indium-oxide, so this is a plausible result.
From Eq.6a and Eq.5 one gets:%

\begin{equation}
\Delta\text{E=E*}(\xi_{\text{0}}/\xi) \tag{6b}%
\end{equation}
that can now be used for an estimate of E*. Using Eq.6b with $\Delta$E$\simeq
$60K for the sample with k$_{\text{F}}\ell$=0.19; $\xi$=23\AA \ (see Fig.9)
gives E*$\approx$120K. Note that, similar to the low values for
E*/E$_{\text{F}}$ obtained in \cite{24} for other systems, the current E*
is\ smaller by more than an order of magnitude than the Fermi-energy of the
material (E$_{\text{F}}$ for In$_{\text{x}}$O with a carrier-concentration
n=8.1$^{\centerdot}$10$^{\text{19}}$cm$^{\text{-3}}$ is $\simeq$1600K).

This value for E* yields $\Delta$E$\simeq$50K for the sample with
k$_{\text{F}}\ell$=0.21; $\xi$=27\AA \ which exhibits diffusive-like
absorption characteristics awhen measured at $\approx$4K. This does not
contradict the observation of an insulating G(T) behavior at this temperature
range; thermal excitation to states lying $\approx$50K above the mobility-edge
will only contribute significantly to the conductance exceeding $\approx$12K.
Note however that the distance to the mobility-edge associated with $\Delta
$E$\simeq$50K is not yet small enough to allow appreciable electron-electron
energy-exchange via virtual transitions to extended states; the time allowed
for diffusion in these states $\hslash$/$\Delta$E is $\approx$10$^{\text{-13}%
}\sec$ which is an order of magnitude shorter than $\gamma_{\text{in}%
}^{\text{e-e}}$ of the diffusive system at these temperatures \cite{5,7}. It
is therefore hard to see how to account for the results in Fig.7 with a
single-particle scenario and with just the implicit assumptions of system
homogeneity made above.

For example one may consider the possibility that a significant part of the
current-carrying-path is composed of regions that are diffusive (and their
combined resistance is comparable to that of the bottleneck resistors). The
occurrence of `metallic-puddles' within the globally insulating system is a
likely scenario when the system approaches the metal-insulator transition from
the insulating side \cite{32}. This could be a consequence of the distributed
nature of the localization-length, a complication that is sometime ignored on
the (misguided) logic that a specific value for $\xi$ is singled-out by
percolation constraints \cite{33}. One should not be surprised to find
deviations from the predictions of these models even for some aspects of the
dc conductivity \cite{34} and anyway, for measurements that involve the bulk
of the sample, a realistic distribution of localization-lengths should be
taken into account. At finite frequencies diffusive regions need not percolate
in the system; it is only necessary that the applied F$_{\text{stress}}$
induces dissipative currents in them thus generating phonons in excess of
those present in equilibrium. This then will be effective in
system-randomization with the ensuing (once Ohmic conditions are restored),
slowly relaxing excess-conductance which contributes to the perceived absorption.

A pertinent consideration here is a conceivable modification of the
wavefunctions due to correlation and many-body effects. The Coulomb
interaction on a scale of $\xi$ derived from the G(T) data is comparable with
$\Delta$E of the samples in Fig.7 therefore hybridization with states above
E$_{\text{C}}$ cannot be ruled out \cite{35}. It is difficult to estimate the
relevance of these processes nearer the transition where the contribution of
electronic polarization to the dielectric-constant becomes dominant \cite{36},
so a self-consistent treatment must be invoked.

Another complication is that the states above E$_{\text{C}}$ are actually
localized in the ground-state; they just \textit{appear} to be extended under
finite temperature or non-Ohmic fields (at high frequencies the latter will be
effective even when the associated conductance-change is smaller than that
required by T due to the relative freedom from percolation constraints). Note
that the many-body density-of-states grows extremely fast with energy (unlike
in a single-particle scenario where this change is algebraic), and
delocalization or an increase of the localization-length \cite{37} may occur
due to the excess energy supplied by the stress-field.

The combination of potential fluctuations, higher-order excitations, and
extended states lying close to the Fermi energy may enhance the occurrence of
metallic "puddles". Obviously, dimensionality plays a role in any of these
scenarios. If however the observed E$_{\text{C}}$ is a many-body mobility-edge
it should also occur in 2D systems albeit probably at a considerably higher
energy. More experiments are needed to elucidate the relative importance of
these mechanisms. The appearance of the T*-problem in so many systems may be
suggestive of an underlying physical effect relevant for the MIT problem. It
clearly deserves to be addressed whether it signifies a real mobility-edge or
it is just a consequence of a finite temperature measurement. The effectively
low\ value of E* (relative to E$_{\text{F}}$) means that, over a wide range of
the physical parameter that is used to characterize the disorder, the system
may be still within the `critical' regime of the metal-insulator transition.

\subsection{Summary}

We have presented results of energy absorption from applied
electromagnetic-fields in three-dimensional In$_{\text{x}}$O samples. For
Anderson-insulating samples that are far from the metal-insulator transition,
the absorption appears to be limited to frequencies that are of the order of
the electron-phonon scattering rates of the material. This suggests that the
hopping process in the range of temperature and disorder studied is mediated
by phonons. Likewise, thermalization of the electronic system then hinges on
the presence of\ a phonon bath. This is in agreement with the conclusions
reached by Fleishman and Anderson \cite{1}. These authors also anticipated a
change of behavior as the mobility-edge is approached from below and expected
that this would be reflected in G(T) that ought to include contribution from
activation to extended states. The experimentally observed change in G(T) as
the mobility-edge is approached appears to be more complicated, and it would
appear that more elaborate models of conductivity need to be developed for the
critical transport regime. These may also shed light on the observation that
absorption from ac fields is more sensitive than temperature to the proximity
of a mobility-edge (whether real or apparent).

The effective range of the interaction, clearly important to these issues, was
not explicitly dealt with in our experiments. At finite temperatures the range
of the Coulomb interaction is limited by the finite conductivity for which the
relevant scale is presumably the hopping-length r(T). At liquid helium
temperatures and for the range of the disorder in the samples used in this
study, r(T) is of order of few hundred angstroms. This is an order of
magnitude larger than the inter-carrier separation n$^{\text{-1/3}}$
suggesting that, for the more disordered samples, our experimental results are
relevant for systems with long-range interaction.

As the disorder in our samples was reduced and the diffusive phase approached,
the absorption extended to progressively higher frequencies. This was observed
on 3D samples but not in 2D samples with comparable degree of disorder. The
similarity of this observation with the T*-problem that appears to be a common
feature in many 3D systems near their MIT was pointed out. It is hoped that
this problem will receive due theoretical attention. Possible relevance of a
nearby mobility-edge for bringing about this behavior as well as the
diffusive-like absorption characteristics of just insulating 3D samples was
discussed. Various effects that might contribute to these phenomena were
mentioned including possible many-body effects. It would be interesting to
extend the absorption study to include the dependence of the absorption on the
amplitude of the stress-field near the transition as it may shed some light on
the relative importance of many-body effects.

\begin{acknowledgement}
This research has been supported by a grant administered by the Israel Academy
for Sciences and Humanities.
\end{acknowledgement}

\end{document}